\documentstyle[11pt,jd24,psfig]{article}
\begin{document}

\title{The frequency content of Double--Mode Cepheids light curves and the
importance of the cross--coupling terms}
\author{E. Poretti}
\affil{Osservatorio Astronomico di Brera, Via E. Bianchi 46\\
22055 Merate, Italy}

\begin{abstract}
The recent results obtained on the frequency content of Double--Mode
Cepheids light curves and the properties of their Fourier parameters
are reviewed. Some points briefly discussed in previous papers are
described.
\end{abstract}

\keywords{Double--mode Cepheids, frequency analysis}

\section{Results of the analysis of DMCs light curves}

Pardo \& Poretti (1997) determined the frequency content of Double--Mode 
Cepheids (DMCs) on the basis of the available $V$ photometry; moreover,
the Fourier decomposition was applied and very accurate sets of Fourier 
parameters were derived. In a second paper, Poretti \& Pardo (1997a) 
discussed their properties by means of generalized phase differences.
The main conclusions of the two  papers were:
\begin{enumerate}
\item Each DMC displays a different content of harmonics and cross--coupling
terms
\item The light curves of Classical Cepheids have the same Fourier parameters
as the Fundamental ($F$) mode of DMCs 
\item The light curves of $s$--Cepheids have the same Fourier parameters
as the First Overtone ($1O$) Mode of DMCs 
\item The discontinuity at $P$=3 d is observed also in the progression of the
Fourier parameters of $1O$ DMC light curves
\item The light curve of the unique Second Overtone ($2O$) pulsator CO Aur
is found to be perfectly sine--shaped (see also Poretti \& Pardo 1997b)
\item No significant third periodicity was found
\item Only the $F$--mode of U TrA shows some hints of period variation
\item Only the $1O$--mode of EW Sct shows some hints of amplitude variation
\item Generalized phase differences have well defined loci in the Fourier
space parameters
\end{enumerate}
Points  2 and 3 constitute an independent confirmation of the different
pulsational nature of Galactic Classical and $s$--Cepheids; the EROS and
MACHO surveys established it for the Cepheids in the Magellanic Clouds on
the basis of the observed Period--Luminosity relationships.

In this contribution we describe several points which could only briefly
be discussed in the previous papers.
\begin{figure}
\centerline{\psfig{figure=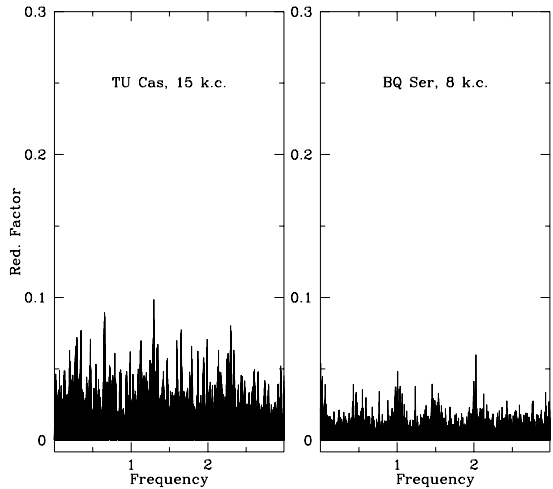,width=8truecm,height=8truecm,rwidth=8truecm,rheight=8truecm,clip=}}
\caption{The last power spectra obtained introducing the known
constituents (k.c.) identified
in the frequency analysis. The presence of
a third frequency was suspected in the cases of TU Cas (left panel)
and BQ Ser
(right panel), but no peak is discernible at the expected $2O$ frequencies (0.825
d$^{-1}$ for TU Cas, 0.42 d$^{-1}$ for BQ Ser). The observed pattern
can be reasonably ascribed to noise effects. }
\end{figure}

\section{No DMC is a triple--mode Cepheid}
Our analysis showed that the DMC light curves can be explained by
the action of two independent frequencies, their harmonics and cross--coupling
terms. In the literature there are two stars for which the presence of a third
independent periodicity was claimed. Figure 1 shows the two last power
 spectra obtained
in the cases of TU Cas and BQ Ser, the two candidates.

In the left panel (TU Cas), the power spectrum obtained by introducing the
15 known components (k.c.'s) is shown;  the highest peak is at 1.296 d$^{-1}$,
but it can be considered just to be a noise effect, looking at the whole
pattern of the peak distribution.
 It should be remarked that the  third frequency (corresponding to
the 2$O$ mode) is expected to be seen around 0.825 d$^{-1}$.

In the case of BQ Ser (right panel), Szabados (1993) found an $f_3$=0.42
d$^{-1}$ 
term; however, the power spectrum obtained by introducing the 8 k.c. shows
a residual noise around the integer values of d$^{-1}$. We often found
this residual noise in our analysis, since it originates in uncertainties
arising from the merging of different sets of data or from instrumental
drifts within
a single set. Regarding the Szabados identification, 
the fact that in his analysis the first harmonic $2f_2$ does not appear
is doubtful, since the $f_2$ amplitude (0.11 mag) and the period (3.3 d)
strongly support the presence of a measurable 2$f_2$ term, i.e.
an asymmetrical light curve.
Moreover, the 0.42 d$^{-1}$ term is very close to the 2$f_2-f_1$
term (0.43 d$^{-1}$) and also all the other terms involving the $f_3$ term
can be easily explained as function of $f_1$ and $f_2$ only: $f_1+f_3=2f_2,
f_2+f_3=3f_2-f_1, f_1+f_2+f_3=3f_1,
f_2+f_3-f_1=3f_2-2f_1$. Hence, very probably the third periodicity does not
exist at all.

We can conclude that after introducing all the
terms detected in the frequency analysis (the two independent frequencies
$f_1$ and $f_2$ and their harmonics and cross--coupling terms) for all
the DMC of our sample the last power
spectrum did not show any other significant terms.

\begin{figure}
\centerline{\psfig{figure=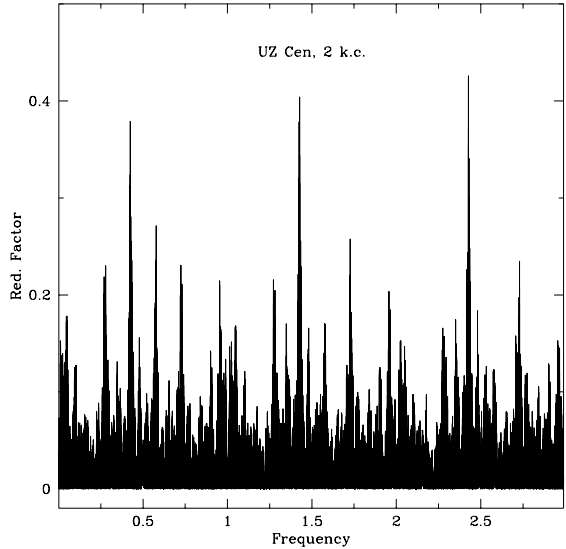,width=8truecm,height=8truecm,rwidth=8truecm,rheight=8truecm,clip=}}
\caption{When searching for the third component in the light curve of
UZ Cen the observed highest peak was not at 0.424 d$^{-1}$, but at
2.427 d$^{-1}$, i.e. an alias. In the subsequent analysis, the 
0.424 d$^{-1}$ value was considered and no residual noise was left at
2.427 d$^{-1}$, clearly indicating that the right choice was done.}
\end{figure}
\begin{figure}
\centerline{\psfig{figure=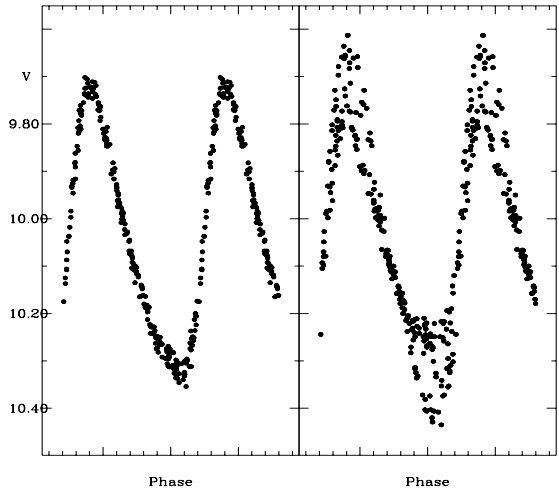,width=8truecm,height=7truecm,rwidth=8truecm,rheight=7truecm,clip=}}
\caption{The cross--coupling terms must be subtracted from the original
measurements to extract the light curves of the two independent frequencies.
In the figure, the case of the $f_1$ term of AP Vel is shown: the light curve
is very regular (left panel), but if the cross--coupling terms are not 
subtracted (right panel), an amplitude modulation appears.}
\end{figure}

\section{The detection of the true peaks}
When identifying a new term in a spectrum, attention should be paid to the
aliasing effects. Indeed, it is always possible that 
a peak corresponding to a small--amplitude term (as a cross--coupling often
is) will be shorter than that of its alias, owing to the interaction with
the noise. Since each set of data is composed of  measurements carried
out in a single site, an $f$ term  produces peaks at $f, 1+f, 1-f,
2+f,2-f, 3-f, ...$ and hence this complicated spectral window has to be
carefully evaluated when interpreting the power spectrum. Moreover,
the terms not yet identified can conspire to move power to a close alias
peak.  Fig. 2 shows the relevant case of the search for the third
component in the light curve of UZ Cen:  the power
spectrum obtained by introducing $f_1, 2f_1$ (i.e. the two terms identified in
the first two steps) shows the highest peak at 2.427 d$^{-1}$,
corresponding to the alias 2+$f_2$. Of course, in the subsequent analysis the
exact value of the second frequency was considered and, as expected, the whole
structure disappeared.

\section{The cross coupling terms}
The light curve of a DMC having a period $P$ cannot be considered
to be the sum of the light curves of a Classical Cepheid having the same
period $P$ and an $s$--Cepheid having an observed period $P_1=0.7~P$.
In a DMC, the action of the cross--coupling terms should not be forgotten:
their excitation produces a distortion of the resulting light curve which
implies cycle--to--cycle variations. As an example of that,
Fig. 3 (right panel) shows the light curve of the first period of 
AP Vel when only $f_2, 2f_2$ were subtracted: the light curve appears to
be noisy, even simulating a modulation of the amplitude at both extrema.
Berdnikov (1992) claimed evidence for such a modulation, but
probably he neglected the cross--coupling term, obtaining a spurious result.
In the left panel of Fig. 3 we subtracted all the cross--coupling term we
found and we obtained a light curve which shows a very regular shape and
constant amplitude. It should be noted that no cross--coupling
terms would be detected if instead of a DMC we were observing a pair
composed of two Cepheids.

Cross--coupling terms are observed in all DMCs, from a minimum of 2
(V367 Sct) to a maximum of 9 (TU Cas). It should be noted that the $f_1+f_2$
and the $f_2-f_1$ terms were always observed and that the amplitude of the
$f_1+f_2$ term is always larger than the $f_2-f_1$ one. Fig. 5 shows the
asymmetrical shape of the $f_1+f_2$ light curve in the case of TU Cas.
Moreover, a third 
 order term can have an amplitude larger than a second order
one.  Hence we cannot recommend truncating the fit to the second
order since a considerable part of the signal might be neglected.

Regarding the harmonic terms (i.e. $2f_1, 3f_1, 4f_1, 2f_2$), a regular 
decrease in amplitude was measured. This should happen every time we
have good phase coverage; if not, it is very probable that the amplitude
of at least one of the harmonics is set to a high value.
It should be also noted that the
$5f_1$ harmonic was never observed, but some fifth order terms arose
($3f_1+2f_2$ and $4f_1+f_2$, TU Cas and U TrA); moreover, 
the $3f_2$ harmonic was never observed, but the third order term
$f_1+2f_2$ was.
\begin{figure}
\centerline{\psfig{figure=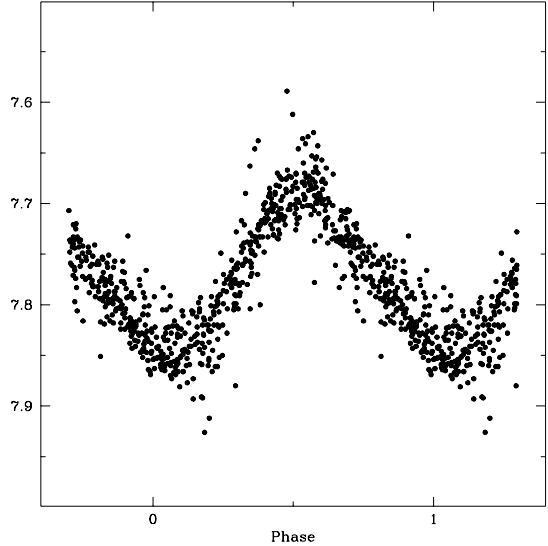,width=8truecm,height=8truecm,rwidth=8truecm,rheight=8truecm,clip=}}
\caption{The frequency content of the TU Cas measurements is very rich; 
the light curve of the cross--coupling term $f_1+f_2$ not only has a
renarkable semi--amplitude (0.078 mag), but its shape is asymmetrical.
Indeed the $2(f_1+f_2)$ term has a semi-amplitude of 0.015 mag.}
\end{figure}
\section{Theoretical models}
The determination of the frequency content of the DMC light curves
provides
a quantitative analysis of the importance of the cross--coupling terms.
The amplitudes of each term is reported by Pardo \& Poretti (1997; Tables
3, 4 and 5).
In our opinion, these results should be used to build more detailed models
of DMC stars, since they are giving clear observational evidence of
the energy distribution in each mode.

\acknowledgments

It is a pleasure to acknowledge fruithful discussions with T.~Aikawa,
E.~Antonello, L.~Mantegazza.

\end{document}